\documentclass[twocolumn,prb,amsmath,amssymb]{revtex4}
\usepackage[T1]{fontenc}
\usepackage[latin1]{inputenc}
\usepackage{graphicx}
\usepackage{amssymb}
\usepackage{dcolumn}
\usepackage{color}
\usepackage{bm}
 
\begin{document}

\title{Electrochemically Top Gated Graphene: Monitoring Dopants by Raman Scattering}

\author{A. Das$^1$}
\author{S. Pisana$^2$}
\author{S. Piscanec$^2$}
\author{B. Chakraborty$^{1}$}
\author{S. K. Saha$^{1}$}
\author{U. V. Waghmare$^3$}
\author{R.Yiang$^{4}$}
\author{H.R.Krishnamurhthy$^1$}
\author{A. K. Geim$^{4}$}
\author{A. C. Ferrari$^2$}
\email{acf26@eng.cam.ac.uk}
\author{A.K. Sood$^{1}$}
\email{asood@physics.iisc.ernet.in}

\affiliation{$^{1}$Department of Physics, Indian Institute of Science, Bangalore 560012,India\\
$^2$Cambridge University, Engineering Department, 9 JJ Thomson Avenue, Cambridge CB3 OFA, UK\\
$^3$ TSU, Jawaharlal Nehru Centre for Advanced Scientific Research,Bangalore 560064, India\\
$^4$ Department of Physics and Astronomy, Manchester
University,UK}

\begin{abstract}
We demonstrate electrochemical top gating of graphene by using a
solid polymer electrolyte. This allows to reach much higher
electron and hole doping than standard back gating. In-situ Raman
measurements monitor the doping. The G peak stiffens and sharpens
for both electron and hole doping, while the 2D peak
shows a different response to holes and electrons. Its position
increases for hole doping, while it softens for high electron
doping. The variation of G peak position is a signature of the
non-adiabatic Kohn anomaly at $\Gamma$. On the other hand, for
visible excitation, the variation of the 2D peak position is ruled
by charge transfer. The intensity ratio of G and 2D peaks
shows a strong dependence on doping, making it a sensitive
parameter to monitor charges.
\end{abstract}

\maketitle

The recent discovery of thermodynamically stable two-dimensional
single and few layers graphene
\cite{NovScience2004,NovNature2005,ZhangNature2005,NovPnas2005} has led to many
experimental and theoretical advances in two dimensional physics and
devices\cite{NovScience2007}. In particular, near ballistic
transport at room temperature and high carrier mobilities (between
3000 and 25000 cm$^{2}$/Vs) \cite{NovNature2005,ZhangNature2005}
make graphene a potential material for nanoelectronics
\cite{Lemme,kimribbon,avouris}.

Electrochemical top gating is key to enable polymer transistors
\cite{henning,dhoot}. It has also been successfully applied for
nanotubes \cite{shim,shim2,liu,forro,rosenblatt}. Here we
demonstrate a top-gated graphene transistor able to span much higher
doping levels than previously reported. Electron and hole doping up
to $\sim$ 5$\times$10$^{13}$cm$^{-2}$ is achieved by solid polymer electrolyte gating. Such a high doping level is
possible because the nanometer thick Debye layer
\cite{liu,SiddonsNanoLett2004,shim2} gives a much
higher gate capacitance compared to the commonly used 300 nm thick
SiO$_{2}$ back gate\cite{geimrev}. A significant advantage of a
solid polymer electrolyte over electrolytes in solution is that it
does not degrade the sample and the electrodes, while the gate leakage
current is negligible compared to the drain current\cite{liu}. Graphene's response to the polymer electrolyte also shows its
potential for both electronic and molecular sensing.

Doping is monitored by in-situ Raman spectroscopy
together with transport measurements. Raman spectroscopy is a
powerful non-destructive technique to identify the number of layers,
structure, doping and disorder
\cite{FerrariPrl2006,SimoneNaturematerials2007,YanPrl2007,SSCReview}. The prominent Raman features in graphene are the G-band at $\Gamma$
($\sim$1584 cm$^{-1}$), and the 2D band at $\sim$ 2700 cm$^{-1}$
involving phonons at $\textbf{K+$\Delta$k}$ points in the Brillouin
zone
\cite{FerrariPrl2006,SSCReview}.
The value of $\textbf{$\Delta$k}$ depends on the excitation laser
energy, due to a double-resonance Raman process and the linear
dispersion of the phonons around K
\cite{ThomsenPrl2000,Piscanec2004,FerrariPrl2006}. The effect of
doping induced by SiO$_{2}$ back gating on the G-band frequency and
full width at half maximum (FWHM) has been reported recently
\cite{SimoneNaturematerials2007,YanPrl2007}. This results in G peak
stiffening and linewidth decrease for both electron and hole doping.
The decrease in linewidth saturates when doping causes a Fermi
level shift bigger than half the phonon energy
\cite{SimoneNaturematerials2007,YanPrl2007}. The strong
electron-phonon coupling in graphene and metallic nanotubes gives
rise to Kohn anomalies in the phonon dispersions
\cite{Piscanec2004,Piscanec2007,Lazzeri2006}, which result in phonon
softening. The G peak stiffening is due to the non-adiabatic removal
of the Kohn-anomaly from $\Gamma$\cite{SimoneNaturematerials2007}.
The FWHM(G) sharpening happens because of the blockage of the decay
channel of phonons into electron-hole pairs due to the Pauli
exclusion principle, when the electron-hole gap becomes higher than
the phonon
energy\cite{SimoneNaturematerials2007,YanPrl2007,Lazzeri2006,LAzzeriPrl2006,ando,castroneto}.
A similar behavior is observed for the LO-G$^{-}$ peak of doped
metallic nanotubes\cite{Wu,shim}, for exactly the same reasons.

In the previous Raman studies on doped graphene
\cite{SimoneNaturematerials2007,YanPrl2007}, a $\sim$300 nm
SiO$_{2}$ gate was used. This limited the maximum doping levels to
less than 1$\times$10$^{13}$cm$^{-2}$. The maximum G peak upshift
was less than 10 cm$^{-1}$. Here, we focus on the simultaneous
evolution of the G and 2D peaks. The G peak stiffens and sharpens
for both electron and hole doping. On the other hand, the 2D peak
shows a different response to holes and electrons. Its position
markedly increases for hole doping, while it softens for high
electron doping. The intensity ratio of 2D and G shows a strong
dependence on doping, making it a sensitive parameter to monitor
Fermi level shifts.

\begin{figure}
\centerline{\includegraphics[width=\columnwidth]{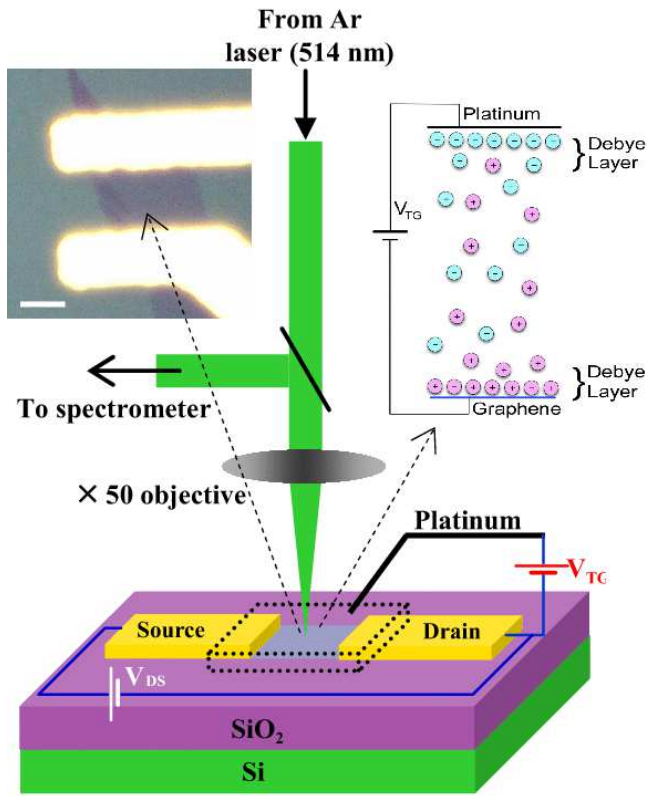}}
\caption{(color online). Schematic diagram of the experimental
setup. The black dotted box between the drain-source indicates the
thin layer of polymer electrolyte (PEO + LiClO$_{4}$). The left
inset shows the optical image of a single layer graphene connected
between source and drain gold electrodes. Scale bar: 5$\mu$m. The
right inset is a schematic illustration of polymer electrolyte top
gating, with Li$^{+}$ and ClO$_{4}^{-}$ ions and the Debye layers
near each electrode.} \label{Figure1}
\end{figure}

Graphene samples are produced by micro-mechanical cleavage of bulk
graphite and deposited on Si covered with 300 nm
SiO$_{2}$ (IDB Technologies LTD)\cite{geimrev}. Raman spectroscopy
is used to select single layers\cite{FerrariPrl2006}. Source and
drain Au electrodes are then deposited by photolithography as shown
in Fig.\ref{Figure1}. Top gating is achieved by using solid polymer
electrolyte consisting of LiClO$_{4}$ and polyethelyne oxide (PEO)
in the ratio 0.12:1, as previously used for nanotubes \cite{liu}.
The gate voltage is applied by placing a platinum electrode in the
polymer layer \cite{liu,SiddonsNanoLett2004}. Electrical
measurements are done using Keithley 2400 source meters. Fig.
\ref{Figure1} shows the schematic of the experimental setup for
transport and Raman measurements. Raman spectra of pristine and back
gated samples are measured with a Renishaw spectrometer. In-situ
measurements on top gated graphene are recorded using a WITEC
confocal (X50 objective) spectrometer with 600 lines/mm grating, 514.5 nm excitation and very low power
level ($\sim$ 1mW) to avoid any heating effect. The spectral resolution of the
two instruments is determined by fitting the Rayleigh line to a
Gaussian profile and is 1.9 cm$^{-1}$ for the Renishaw
spectrometer and 9.4 cm$^{-1}$ for WITEC spectrometer. The Raman
spectra are then fitted with Voigt functions. The FWHM of the
Lorentzian components give the relevant information on the phonon
lifetime. Note that a very thin layer of polymer electrolyte does
not absorb the incident laser light. Furthermore the Raman spectrum
of the polymer does not cover the signatures of graphene, as will be
discussed later. The measured  source-drain currents (I$_{SD}$) and G, 2D peaks
are reversible at different gate voltages. In transport experiments
a small hysteresis in current ($\sim$1 $\mu$A) is observed during
forward and backward gate voltage scans (at a intervals of 10 minutes for each
gate voltage step). On the other hand, the Raman hysteresis $\leq$ 1 cm$^{-1}$.

We finally compare our experimental results with Density Functional
Perturbation theory (DFPT) simulations\cite{Baroni2001}.
Calculations are performed within the generalized gradient
approximation (GGA) \cite{Perdew1996}. We use plane-waves
($30~{\rm Ry}$ cut-off) and pseudopotential~ \cite{Troullier1991}
approaches.
The semi-metallic character of the system is treated by performing
the electronic integration with a Fermi-Dirac first-order
spreading with a smearing of 0.01 Ry \cite{Methfessel1989}.
Integration over the BZ is done with an uniform
$72\times72\times1$ k-points grid.
Calculations are done using the Quantum Espresso code~\cite{PWscf}.

    We first consider the electrical response. In
order to compare our top gating results with back gating
measurements, it is necessary to convert the top gate voltage into
an effective doping concentration. In general, the application of a
gate voltage (V$_{G}$) creates an electrostatic potential difference
$\phi$ between graphene and the gate electrode and a Fermi level, E$_{F}$, shift as a result of
addition of charge carriers. Therefore,
\begin{equation}
V{_{G}}=\frac{E{_{F}}}{e}+\phi
\end{equation}
$\phi$ and E$_{F}$/e are determined by the geometrical capacitance,
$C{_{G}}$, and the chemical (quantum) capacitance of graphene,
respectively.

    Let us first consider back gating (BG). For a back gate, $\phi$ =
$\frac{ne}{C{_{BG}}}$ where n is the carrier concentration and
C$_{BG}$ is the geometrical capacitance. For single layer graphene
$C{_{BG}}=\frac{\epsilon\epsilon{_{0}}}{d_{BG}}$, where $\epsilon$
is the dielectric constant of SiO$_{2}$ ($\sim$4), $\epsilon_{0}$ is
the permitivity of free space and d$_{BG}$ is 300 nm. This results
in a very low gate capacitance C$_{BG}$ =
1.2$\times$10$^{-8}$Fcm$^{-2}$. Therefore, for a typical value of n
= 1$\times$10$^{13}$ cm$^{-2}$, the potential drop is $\phi$ = 100
V, much larger than $\frac{E{_{F}}}{e}$. Hence, $V{_{BG}}$ $\approx$
$\phi$ and the doping concentration becomes n = $\eta$V$_{BG}$,
where $\eta$ = C$_{BG}$/e. However, most samples have a zero-bias
(V$_{BG}$=0) doping of, typically, a few 10$^{11}$
cm$^{-2}$\cite{NovScience2004,geimrev,cinzin}. This is reflected in
the existence of a finite gate voltage V$_{nBG}$ at which the Hall
resistance is zero  and the longitudinal resistivity reaches its
maximum. This maximum is associated with the Fermi level positioned between the valence and the conduction bands (the Dirac point). Accordingly, a positive
(negative) V$_{BG}$-V$_{nBG}$ induces electron (holes) doping,
with an excess-electron surface-concentration of n=$\eta$(V$_{BG}$
- V$_{nBG}$). A value of
$\eta\approx$7.2$\times$10$^{10}$cm$^{-2}V^{-1}$ is found from Hall
effect measurements and agrees with the estimation from the gate
geometry\cite{NovNature2005,NovScience2004,ZhangNature2005,NovPnas2005}.

\begin{figure}
\centerline{\includegraphics[width=\columnwidth]{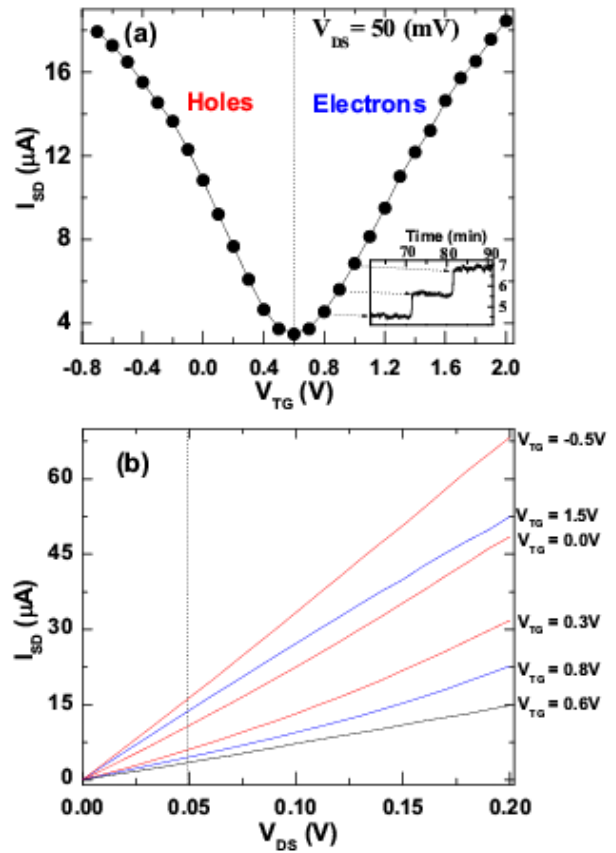}}
\caption{(color online). (a) I$_{SD}$ as a function of top gate
voltages (V$_{TG}$). The inset shows the I$_{SD}$ time dependence
at fixed V$_{TG}$. The dotted line corresponds to the Dirac point.
(b) I$_{SD}$ vs V$_{DS}$ at different top gate voltages. Red and
blue lines correspond to hole and electron doping, respectively.
The black dotted line corresponds to the source-drain value at
which the gate dependence curve is measured.} \label{Figure2}
\end{figure}

    Let us consider the present case of top gating (TG). First we briefly discuss how the polymer electrolyte
works as a gate. When a field is applied, free cations tend to
accumulate near the negative electrode, creating a positive charge
there and an uncompensated negative charge near the interface. The accumulation is limited by the concentration
gradient, which opposes the Coulombic force of the electric field.
When a steady state is reached, the statistical space charge
distribution resembles that shown in Fig. 1. This layer of charge
around an electrode is called the Debye layer. As shown in Fig.
\ref{Figure1}, when we apply a positive potential (V$_{TG}$) to the
platinum top gate, with respect to the source electrode connected to graphene,
the Li$^{+}$ ions become dominant in the Debye layer formed
at the interface between graphene and the electrolyte. The Debye layer of thickness d$_{TG}$ acts like a parallel
plate capacitor. Therefore, the geometrical capacitance in this case
is $C{_{TG}} = \frac{\epsilon\epsilon{_{0}}}{d_{TG}}$, where
$\epsilon$ is the dielectric constant of the PEO matrix. The Debye
length is given by d$_{TG}$ =
(2ce$^{2}$/$\epsilon$$\epsilon$$_{0}$kT)$^{-1/2}$ for a monovalent
electrolyte where c is the concentration of the electrolyte, e is
the electric charge, kT is the thermal energy \cite{rafael}. In
principle, d$_{TG}$ can be calculated if the electrolyte
concentration is known. However, in presence of a polymer, the
electrolyte ions form complexes with the polymer
chains~\cite{RogerMacromolecules1999,SalomanJPC1994}. Hence the
exact concentration of ions is not amenable to measurement. For
polymer electrolyte gating the Debye layer thickness is reported
to be a few nanometers (1 $\sim$ 5 nm)
\cite{liu,DonleyMacromocules1997}. The dielectric constant
($\epsilon$) of PEO is 5 \cite{DielecPEO}. Assuming a Debye length
of 2 nm, we get a gate capacitance C$_{TG}$ = 2.2$\times$ 10$^{-6}$
F cm$^{-2}$, which is much higher than C$_{BG}$. Therefore, the first
term in equation (1) cannot be neglected. The Fermi energy in
graphene changes as
$E{_{F}}(n)={\hbar}\left|{v{_{F}}}\right|\sqrt{{\pi}n}$, where
$\left|{v{_{F}}}\right|$ = 1.1 $\times$
10$^{6}m s^{-1}$\cite{NovNature2005,ZhangNature2005,YanPrl2007} is the
Fermi velocity, hence:
\begin{equation}
V{_{TG}}=\frac{{\hbar}\left|{v{_{F}}}\right|\sqrt{{\pi}n}}{e}+\frac{ne}{C{_{TG}}}
\end{equation}
Using the values of C$_{TG}$ and  $v{_{F}}$,
\begin{equation}
V{_{TG}}(volts)=1.16\times10^{-7}\sqrt{n}+0.723\times10^{-13}n
\end{equation}
where n is in units of cm$^{-2}$. Eq. 3 allows us to estimate the
doping concentration at each top gate voltage (V$_{TG}$). Note
that, as in back gating, we also get the minimum source-drain
current at finite top gate voltage (V$_{nTG}$ = 0.6 V), as seen in
Fig. \ref{Figure2}a. Accordingly, a positive (negative)
V$_{TG}$-V$_{nTG}$ induces electron (holes) doping.

    Fig. \ref{Figure2}a shows the source-drain current (I$_{SD}$) of the top
gated graphene as a function of electrochemical gate voltage. Note
that for each point a given gate voltage is applied for 10 minutes
to stabilize I$_{SD}$. The gate dependence of the drain current
(Fig. \ref{Figure2}a) shows ambipolar behavior and is almost
symmetric for both electron and hole doping. This directly relates
to the band structure of graphene, where both electron and hole
conduction are accessible by shifting the Fermi level. The
I$_{SD}$-V$_{SD}$ characteristics at different electrochemical gate
voltages (Fig. \ref{Figure2}b) show linear behavior, indicating the
lack of significant Schottky barriers at the electrode-graphene
interface.

\begin{figure}
\centerline{\includegraphics[width=60mm]{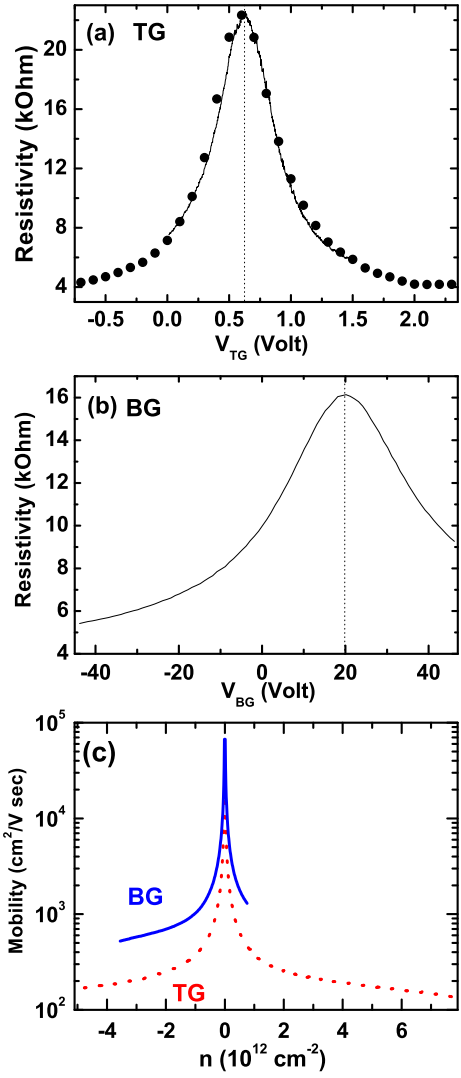}}
\caption{(Color online).(a) Resistivity as a function of top
gate. The dots are extracted from Fig.\ref{Figure2}a
for W/L = 1.55. The solid line corresponds to the resistivity
change as a function of V$_{TG}$, where the gate voltage is varied
at a intervals of 2mV. (b) Resistivity of the same sample as a
function of back gate. The dotted black line marks the
Dirac point. (c) Mobility as function of doping for TG (red
dotted) and (BG) (solid blue).} \label{Figure3}
\end{figure}

    Fig. \ref{Figure3}a plots the resistivity of our graphene layer
(extracted from Fig. \ref{Figure2}a knowing the sample's aspect
ratio: W/L = 1.55) as a function V$_{TG}$. Fig.\ref{Figure3}b shows
the back gate response of the same sample (without electrolyte). There is an increase in resistivity maximum ($\sim$ 6
k$\Omega$) after pouring the electrolyte, which may
originate from the creation of more charged impurities on the
sample. Fig. \ref{Figure3} also shows that for both TG and BG
experiments the resistivity does not decay sharply around the Dirac
point. Indeed, it has been suggested that the sharpness of the resistivity
around the Dirac point and the finite offset gate voltage
(V$_{nBG}$) depend on charged impurities
\cite{KimCondmat,SarmaCondmat,geimrev}.

The conductivity minimum ($\sigma$$_{min}$) (resistivity maximum) is
obtained when the Fermi level is at the Dirac point. This is
generally around $\sim$ 4$\frac{e^{2}}{h}$
\cite{NovNature2005,geimrev}. In both our back and top gate experiments
the conductivity minimum is reduced by the contact resistance, since measurements are performed in the two-probe configuration. As the contact resistance is expected to depend strongly on the carrier concentration (due to formation of a p-n junction around the contact and changing in the density of states in graphene) we can only give an estimation of the lower bound of the contact resistance. As shown in
Fig. \ref{Figure3}a, the resistivity saturates at $\sim$ 4.2
k$\Omega$ (which corresponds to a 2.7 k$\Omega$ resistance) when the
carrier density in the sample is the highest. Therefore, we estimate our contact resistance to be around 2.7 k$\Omega$. Subtracting
the effect of contact resistance, the minimum
conductivity in our sample is $\sim1.3\frac{e^{2}}{h}$. Minimum
conductivities in the range from 2$\frac{e^{2}}{h}$ to
10$\frac{e^{2}}{h}$ were recently reported\cite{KimCondmat}, with
the spread assigned to charged impurities.

Fig.\ref{Figure3}c shows the change in mobility (using the simple
Drude model $\mu$ = (en$\rho$)$^{-1}$\cite{KimCondmat}) as a
function of doping for our TG/BG experiments. The
mobility is smaller in the TG case. This is consistent with the
reduction in conductivity minimum and can be attributed to the
presence of added charge impurities from the polymer electrolyte.
Despite the limitations in `on' and `off' currents, our large
graphene device shows an on/off ratio of $\sim$ 5.5. This is higher
than what previously reported for devices using 20 nm thick
SiO$_{2}$ as top gate (on/off ratio $\sim$ 1.5)\cite{Lemme} and 40
nm thick PMMA as a top gate (on/off ratio $\sim$
2)\cite{GordonCondmat}.

\begin{figure}
\centerline{\includegraphics[width=70mm]{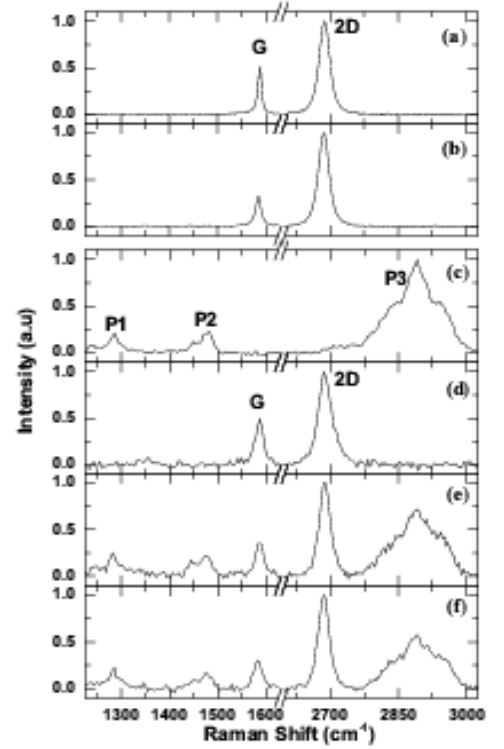}}
\caption{Raman spectra at (a) V$_{BG}$ = 0.0 V and (b)
V$_{BG}$=V$_{nBG}$=20V. (c) Raman spectra of PEO+LiClO$_{4}$
mixture. (d) Raman spectra of graphene before pouring the polymer
electrolyte. (e) Raman spectra at V$_{TG}$ = 0.0 V and (f)
V$_{TG}$=V$_{nTG}$=0.6V. P1, P2 and P3 are the polymer
peaks\cite{Poly}} \label{Figure4}
\end{figure}

We now consider the evolution of the Raman spectra. Fig.
\ref{Figure4}a,b plot the Raman spectra for
V$_{BG}$ = 0 V and V$_{nBG}$ = 20 V. Fig. \ref{Figure4}c to
\ref{Figure4}f show the spectra recorded during the top gate
experiment. Fig. \ref{Figure4}c is the PEO Raman spectrum. This has
three prominent peaks at $\sim$ 1282 cm$^{-1}$ (P1), 1476 cm$^{-1}$
(P2) and 2890 cm$^{-1}$ (P3), which correspond to twisting, bending
and stretching modes of the CH$_{2}$ bonds in the polymer\cite{Poly}. Luckily they do not overlap the main features of
graphene (see Figs. \ref{Figure4}d,f). Furthermore
these PEO Raman lines do not change with gating. Table 1 shows the
comparison of G peak position, Pos(G), FWHM(G) and 2D/G height
ratio, I(2D)/I(G), at zero gate voltage and the Dirac point for BG and
TG.

\begin{table}[h!t!p!]
\caption{G peak position, FWHM and 2D/G height ratio.}
\begin{center}
\begin{tabular}{|c|c|c|c|c|}
\hline
Gate Voltage & Pos(G) & FWHM(G) & I(2D)/I(G) \\
& (cm$^{-1}$) & (cm$^{-1}$) & (height ratio)\\
\hline
V$_{BG}$ = 0.0 V & 1586.7 & 8.7 & 2.0 \\
\hline
V$_{nBG}$ = 20 V & 1584.0 & 12.6 & 3.1 \\
\hline
V$_{TG}$ = 0.0 V & 1586.4 & 13.9 & 2.75 \\
\hline
V$_{nTG}$ = 0.6 V & 1583.1 & 14.9 & 3.3 \\
\hline
\end{tabular}
\end{center}
\label{table1}
\end{table}

Table 1 shows that at the Dirac point (V$_{nBG}$ and
V$_{nTG}$) we have the lowest Pos(G), maximum FWHM(G) and I(2D)/I(G). However, our sample has a lower I(2D)/I(G), FWHM(G) and
higher Pos(G) than the most intrinsic samples measured to
date\cite{cinzin,FerrariPrl2006}, due to the presence of charge
impurities\cite{cinzin}.

\begin{figure}
\centerline{\includegraphics[width=70mm]{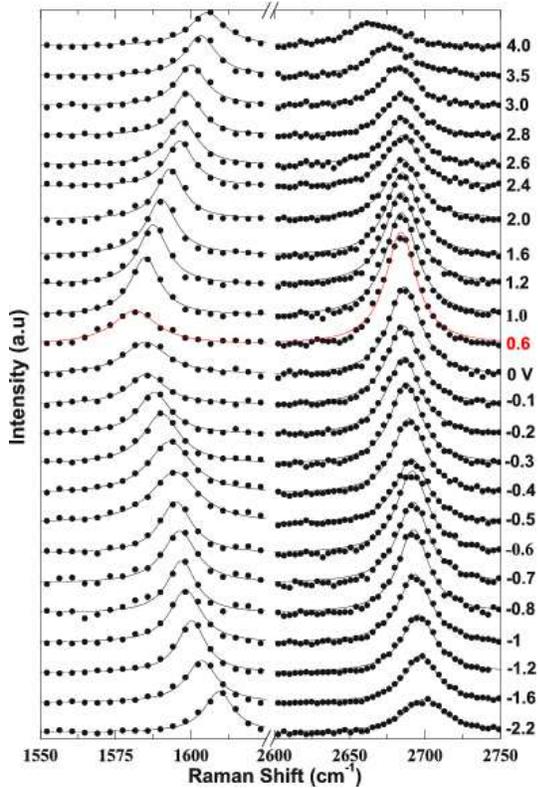}}
\caption{(Color online). Raman spectra at several
V$_{TG}$. The dots are the experimental data. Black lines are
fitted Lorentzians. The red line corresponds to the Dirac point.}
\label{Figure5}
\end{figure}

\begin{figure}
\centerline{\includegraphics[width=60mm]{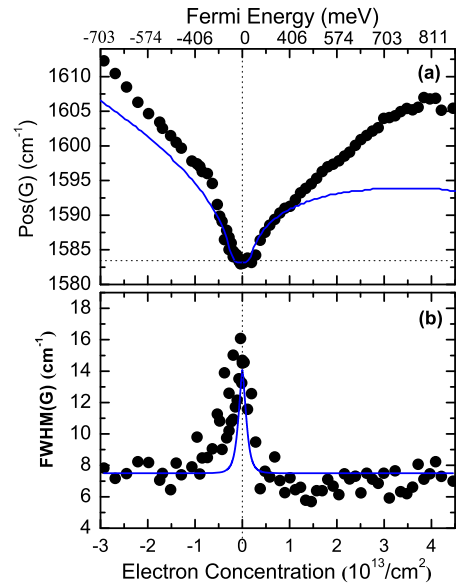}}
\caption{(color online). (a) Pos(G) and (b) FWHM(G) as a function
of electron and hole doping. The solid blue lines are the
predicted non-adiabatic trends from
Refs.\cite{LAzzeriPrl2006,SimoneNaturematerials2007}.}
\label{Figure6}
\end{figure}

\begin{figure}
\centerline{\includegraphics[width=60mm]{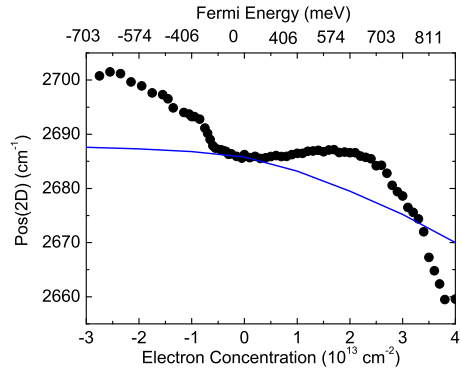}}
\caption{(color online).(a) Pos(2D) as a function of doping. The solid line is our adiabatic DFT
calculation.} \label{Figure7}
\end{figure}

Fig.\ref{Figure5} plots the Raman spectra as
a function of top gate voltages. Fig. \ref{Figure6},\ref{Figure7}
show the Raman parameters as a function of doping. The minimum Pos(G) ($\sim$ 1583.1 cm$^{-1}$) is at V$_{TG}$ = V$_{nTG}$ $\sim$
0.6 V. Pos(G) increases for positive (V$_{TG}$ - V$_{nTG}$) and
negative (V$_{TG}$ - V$_{nTG}$), i.e. for both electron and hole
doping, by up to 30 cm$^{-1}$ for hole
doping and 25 cm$^{-1}$ for electron doping (see Fig.
\ref{Figure6}a). The decrease in FWHM(G) (see Fig. \ref{Figure6}b) for both
hole and electron doping is similar to earlier results
\cite{SimoneNaturematerials2007,YanPrl2007}, even though extended to a much wider doping range. Most interestingly, the
2D peak shows a very different dependence on gate voltages when
compared to the G mode. For electron doping, Pos(2D) does not change
much ($<$ 1 cm$^{-1}$) until gate voltage of $\sim$ 3V
(corresponding to $\sim3.2\times$10$^{13}$cm$^{-2}$). At higher gate
voltages, there is a significant softening by $\sim$ 20 cm$^{-1}$
and for hole doping, Pos(2D) increases by $\sim$ 20 cm$^{-1}$ (see
Fig. \ref{Figure7}a).

\begin{figure}
\centerline{\includegraphics[width=60mm]{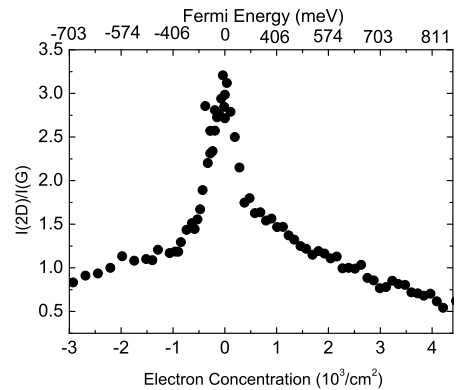}}
\caption{I(2D)/I(G) as a function of hole and electron doping.}
\label{Figure8}
\end{figure}

Fig.\ref{Figure8} plots the variation of G and 2D intensity ratio (I(2D)/I(G))as a function of doping. These show a strong
dependence on doping. The dependence of the 2D mode is much stronger
than that of the G mode and hence the 2D/G intensity ratio is a strong function of the gate voltage. Therefore,
this ratio is an important parameter to estimate the doping density. A similar dependence is
expected for nanotubes.
Figs.\ref{Figure8},\ref{Figure6} also show that I(2D)/I(G) and G peak position should not be used to
estimate the number of graphene layers, unlike what suggested in refs.\cite{GuptaNanoLett2006, GrafNanoLett2007}. The shape of the 2D being the most effective way to identify a single layer\cite{FerrariPrl2006}.

The theoretical trends in Fig.6 were discussed before
\cite{LAzzeriPrl2006,SimoneNaturematerials2007}. These confirm
previous back gate experiments, but extend the data to a much wider
electron and hole range\cite{SimoneNaturematerials2007}. In this
wider range, the theory still captures the main features, such as the
asymmetry between electrons and hole doping\cite{LAzzeriPrl2006},
however the quantitative agreement is poor for large doping, and
requires to reconsider the non-adiabatic calculations of
Ref.\cite{LAzzeriPrl2006}.

Here we focus on the novel trend of 2D peak position as a function
of doping. \textit{This is experimentally and conceptually different from
the G peak}.

The 2D peak is due to second order, double resonant (DR)
Raman scattering~\cite{ThomsenPrl2000,FerrariPrl2006,reich2001}.
In this process, the incoming laser radiation creates an
electron-hole pair close to the Fermi point
$\textbf{k}_F=\textbf{K}$.
The photo-excited electron is then scattered towards the second
inequivalent Fermi point $\textbf{k}_F=\textbf{K}'$ by a phonon of
energy $\hbar\omega_q$ and wavevector \textbf{q}.
A scattering event with a second phonon, of the same energy but
opposite momentum, brings the
electron back to its original position in reciprocal space.
The recombination of the electron-hole couple finally results in
the emission of a photon, whose energy is decreased by
$2\hbar\omega_{{\bf q}}$ with respect to the incoming laser
radiation.
The order of these four events is not fixed, and all their
combinations are possible and have to be taken into account
\cite{reich2001}.

The position of the 2D-peak can be evaluated by
computing the energy of the phonon involved in the second-order,
double resonant scattering process.
As shown in Ref.~\cite{FerrariPrl2006}, due to the trigonal
warping of the $\pi-\pi^*$ bands and the angular dependence of the
electron-phonon coupling (EPC), only phonons
oriented along the ${\bf \Gamma{\bf K{\bf M}}}$ direction and with
${\bf q>{\bf K}}$ give a non-negligible contribution to the
2D-peak.
The precise value of $q$ is fixed by the constraint that the
energy of the incoming photons $\hbar\omega_{L}$ has to exactly
match a real electronic transition.
In particular only a wavevector $\textbf{q}'$ can be found for
which $\hbar\omega_{L}=\epsilon(\pi^{*},{\bf q')-\epsilon(\pi,{\bf
q')}}$, where $\epsilon(n,{\bf k)}$ is the energy of an electron
of band index $n$ and wavevector \textbf{k}, and $\textbf{q}'$ is
measured from \textbf{K} and is in the ${\bf \Gamma{\bf K{\bf
M}}}$ direction.
Once $\textbf{q}'$ has been determined, $q=2q'+K$.
Among the six phonons corresponding to the $\textbf{q}$ vector
that satisfy the DR conditions, only the highest optical branch
has an energy compatible to the measured Raman shift.
Therefore, the theoretical position of the 2D peak corresponds to
twice the energy of the Raman active phonon.

In order to compare with our experiments performed at 514nm, we consider $\hbar\omega_{L}=2.5$ eV. Assuming the
$\pi/\pi^{*}$ bands to be linear, with a slope of 14.1 eV
\cite{Piscanec2004}, this laser energy selects a phonon with
wavevector \textbf{q} of modulus 0.844 in $\frac{2\pi}{a_{0}}$
units, $a_{0}$ being the lattice parameter of graphene.
The dependence of Pos(2D) on doping can be investigated by
calculating, within a DFT framework, the effects of the Fermi
level shift on the phonon frequencies.

In doped graphene, the Fermi energy shift induced by doping
gives two major effects: (i) a change of the
equilibrium lattice parameter with a consequent
stiffening/softening of the phonons, and (ii) the onset of effects
beyond the Adiabatic Born-Oppenheimer approximation that modify
the phonons close to the Kohn
anomalies\cite{SimoneNaturematerials2007,LAzzeriPrl2006}.
The excess (defect) charge gives an expansion (contraction)
of the crystal lattice. This was extensively investigated
to understand graphite intercalation compounds
\cite{Pietronero1981}.
We model the Fermi surface shift by varying the number
of electrons in the system. Since the total energy of charged systems diverges, electrical
neutrality is achieved by imposing a uniformly charged background.
To avoid electrostatic interactions between graphene and the background, the equilibrium lattice
parameter of the charged system is computed in the limit of a
infinite volume unit cell.
Such limit is reached using a model with periodic
boundary conditions where the graphene layers are spaced by 60~\AA.
Phonon calculations for charged graphene are done using the same
unit cells employed for the determination of the corresponding
lattice parameter.
Interestingly, while we observe that for charged graphene the
frequency of zone boundary TO phonons converge only for layer
spacing as large as 60~\AA, the frequency of the $E_{2g}$ mode is
already converged for a 7.5~\AA~spacing.

Dynamic effects beyond Born-Oppenheimer play a
fundamental role in the description of the KA in single wall
carbon nanotubes and in graphene
\cite{Piscanec2007,LAzzeriPrl2006,SimoneNaturematerials2007}.
However, for the 2D peak measured at 514nm the influence of
dynamic effects is expected to be negligible, since the phonons
giving rise to the 2D-peak are far away from the Kohn anomaly at
\textbf{K}.
Thus, we can calculate the position of the 2D-peak without dynamic corrections.

The comparison between the theoretical and the experimental position
of the 2D peak is shown in Fig. \ref{Figure7}a by a solid line.
Our calculations are in qualitative agreement with experiments, considering the spectral resolution and the Debye layer estimation.
Indeed, as experimentally determined, the position of the 2D peak is
predicted to decrease for an increasing electron concentration in
the system. This allows to use the 2D peak to discriminate between electron and hole doping.

The trade-off between measured and theoretical data can be partially explained in terms of
the electrostatic difference existing between the experiments and the model DFT system.
In our simulations the 2D phonon frequencies are very sensitive to
the charged background used to ensure global electrical
neutrality. In the experiments the electric charge on the graphene
surface is induced by capacitative coupling. The electrostatic
interaction between graphene and the electrolyte could thus
further modify the 2D phonons. This does not affect the G peak to
the same extent, due to the much lower sensitivity the G phonon to
an external electrostatic potential.

In conclusion, we have demonstrated graphene top gating using a
solid polymer electrolyte. We reached much higher electron and hole
doping than standard SiO$_{2}$ back gating. The conductivity minimum
and mobility are reduced due to presence of charges. The
Raman measurements show that the G and 2D peaks have different
doping dependence and the 2D/G height ratio changes significantly with
doping, making Raman an ideal tool for graphene nanoelectronics.

We thank K. S. Novoselov for useful discussions. AKS thanks the Department of Science and Technology, India for the
financial support. SP acknowledges funding from Pembroke College and the Maudslay Society. ACF from the Royal Society
and Leverhulme Trust.

\end{document}